# 4D-ACFNet: A 4D Attention Mechanism-Based Prognostic Framework for Colorectal Cancer Liver Metastasis Integrating Multimodal Spatiotemporal Features

Zesheng Li†
School of Computer Science and Technology
University of Chinese Academy of Sciences
Beijing, China
2022111515@stu.sufe.edu.cn

Wei Yang†
School of Computer Science and Technology
University of Chinese Academy of Sciences
Beijing, China
sheepwei_2023@qq.com

Yan su
School of Computer Science and Technology
University of Chinese Academy of Sciences
Beijing, China
1227698971@qq.com

Yiran Zhu
School of Computer Science and Technology
University of Chinese Academy of Sciences
Beijing, China
ciaran_study@yeah.net

Yuhan Tang
School of Computer Science and Technology
University of Chinese Academy of Sciences
Beijing, China
yuhantang012@163.com

Haoran Chen
School of Computer Science and Technology
University of Chinese Academy of Sciences
Beijing, China
202308611@stu.sicau.edu.cn

Chengchang Pan*
School of Computer Science and Technology
University of Chinese Academy of Sciences
Beijing, China
chpan.infante@qq.com

Honggang Qi*
School of Computer Science and Technology
University of Chinese Academy of Sciences
Beijing, China
hgqi@ucas.ac.cn

## ABSTRACT

Postoperative prognostic prediction for colorectal cancer liver metastasis (CRLM) remains challenging due to tumor heterogeneity, dynamic evolution of the hepatic microenvironment, and insufficient multimodal data fusion. To address these issues, we propose 4D-ACFNet, the first framework that synergistically integrates lightweight spatiotemporal modeling, cross-modal dynamic calibration, and personalized temporal prediction within a unified architecture. Specifically, it incorporates a novel 4D spatiotemporal attention mechanism, which employs spatiotemporal separable convolution (reducing parameter count by 41%) and virtual timestamp encoding to model the interannual evolution patterns of postoperative dynamic processes, such as liver regeneration and steatosis. For cross-modal feature alignment, Transformer layers are integrated to jointly optimize modality alignment loss and disentanglement loss, effectively suppressing scale mismatch and redundant interference in clinical-imaging data. Additionally, we design a dynamic prognostic decision module that generates personalized interannual recurrence risk heatmaps through temporal upsampling and a gated classification head, overcoming the limitations of traditional methods in temporal dynamic modeling and cross-modal alignment. Experiments on 197 CRLM patients demonstrate that the model achieves 100% temporal adjacency accuracy (TAA), with performance significantly surpassing existing approaches. This study establishes the first spatiotemporal modeling paradigm for postoperative dynamic monitoring of CRLM. The proposed framework can be extended to prognostic analysis of multi-cancer metastases, advancing precision surgery from "spatial resection" to "spatiotemporal cure."

## 1 Introduction

Colorectal cancer (CRC) ranks as the third most common malignancy globally and the second leading cause of cancer-related mortality [1, 2]. Approximately 50% of CRC patients develop colorectal liver metastasis (CRLM), and hepatic resection combined with neoadjuvant chemotherapy remains the only potentially curative treatment for long-term survival [3, 4]. However, even after radical surgery, 60–70% of patients experience treatment failure due to postoperative recurrence, with a 5-year overall survival rate below 30% [4, 5]. This critical clinical challenge stems from the high heterogeneity of CRLM: differences in gene expression profiles between primary and metastatic lesions (e.g., KRAS mutations, consensus molecular subtypes [CMS]) significantly influence chemotherapy sensitivity and recurrence patterns [5, 6]. Concurrently, pathological alterations in non-tumorous liver tissue (e.g., chemotherapy-induced steatosis, sinusoidal dilatation) regulate tumor cell invasiveness and vessel co-option through inflammatory factors such as TGFβ1 [5, 7]. Moreover, postoperative recurrence risk follows nonlinear temporal evolution, which conventional static models (e.g., the Fong Clinical Risk Score) fail to capture, particularly in tracking dynamic postoperative processes like liver regeneration and steatosis [8, 9].

Recent advances in deep learning-based medical image analysis have improved tumor prognosis prediction, yet existing models exhibit notable limitations. Radiomics quantifies tumor heterogeneity by extracting morphological and textural features from CT images but lacks sensitivity to microenvironmental dynamics such as hepatic steatosis [10]. Clinical models (e.g., carcinoembryonic antigen CEA combined with methylated SEPT9) enhance specificity for metastasis detection (AUC = 0.85) but suffer from low spatial resolution, hindering precise localization of recurrence-prone regions [11]. Furthermore, most deep learning models rely on preoperative static imaging data, neglecting the temporal effects of postoperative liver regeneration and microenvironment remodeling, thereby limiting their capacity to model dynamic risk trajectories [5].

To address these challenges, we propose an innovative multimodal spatiotemporal joint modeling framework. By leveraging a lightweight 4D spatiotemporal attention mechanism, our approach decouples the interannual dynamic evolution of postoperative liver regeneration and steatosis, overcoming the inherent limitations of traditional static models in temporal modeling while reducing parameter count by 41%. Additionally, we design a Transformer-based cross-modal feature alignment strategy that jointly optimizes modality alignment and disentanglement losses, effectively mitigating scale mismatch and redundant interference in clinical-imaging data. Furthermore, a dynamic prognostic decision module is introduced, integrating temporal upsampling and a gated classification head to generate personalized interannual recurrence risk heatmaps. This enables dynamic visualization of spatiotemporal risk evolution, establishing the first interpretable spatiotemporal decision paradigm for postoperative CRLM monitoring.

Compared to existing methods, our primary contributions are summarized as follows: 1) We propose a novel architecture employing spatiotemporal separable convolution and temporal encoders to simulate interannual postoperative changes in hepatic structures, enhancing the model's sensitivity to vessel co-option and steatosis dynamics; 2) A disentanglement loss function is introduced to align radiomic features with clinical indicators, suppressing redundant intermodal information interference in prognosis prediction; 3) By integrating temporal upsampling and a gated classification head, our framework generates patient-specific interannual recurrence risk heatmaps, aiding clinicians in formulating dynamic postoperative surveillance strategies.

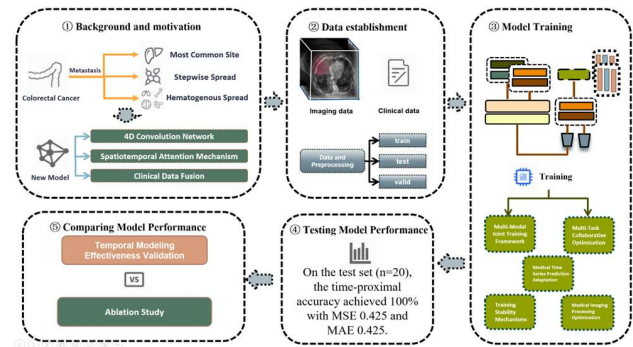

**Figure 1: In this experiment, the motivation is determined through research, the data set is constructed and the model is trained, and finally the superiority of the method is verified through evaluation and experiment.**

## 2 Related Work

In the field of colorectal cancer liver metastasis (CRLM) prognosis research, traditional prognostic models have long been constrained by their inherent limitations. These models predominantly rely on clinicopathological characteristics (e.g., CEA levels, number of liver metastases) and static imaging assessments but fail to incorporate dynamic modeling capabilities. As highlighted by Brudvik et al. [12] in their systematic review, conventional models based on clinicopathological factors struggle to achieve precise predictions due to their neglect of the dynamic evolution of tumor biology and nonlinear associations between genetic mutations and prognosis. Furthermore, widely adopted Cox proportional hazards models [13], grounded in linear risk assumptions, cannot characterize the temporal interactions between postoperative liver regeneration and neoadjuvant chemotherapy[14]. These shortcomings have motivated researchers to explore novel methodologies for improving prognostic accuracy.

Recent advances in integrating CT/MRI radiomics with deep learning techniques have provided new avenues for quantifying tumor heterogeneity. For instance, Tan et al. [15] developed a multimodal deep learning framework that enables tumor heterogeneity quantification without requiring manual segmentation. Zhou et al. [16] extended the application of 3D convolutional networks by proposing an integrated framework based on multiscale feature extraction. By leveraging radiomic data from CT/MRI to construct prognostic models and employing adaptive receptive field adjustment strategies to enhance spatial resolution, their approach achieved improved accuracy in assessing tumor heterogeneity and microenvironmental changes [16]. However, most existing methods focus on preoperative static imaging features and inadequately model the spatiotemporal correlations of postoperative liver regeneration and dynamic microenvironment remodeling.

Subsequently, spatiotemporal dynamic modeling has gained attention in medical imaging, yet its applicability to CRLM remains challenging. Guo et al. [19] proposed a trimodal temporal fusion framework validated in surgical scene graph generation tasks, demonstrating the efficacy of dynamic modeling. Bastiancich et al. [18] utilized single-photon emission computed tomography (SPECT) imaging to evaluate blood-brain barrier permeability and analyzed tumor microenvironment complexity through multiparametric flow cytometry and two-photon imaging [18]. However, these methods, not optimized for CRLM-specific biological processes, are difficult to directly apply to prognosis prediction. To address these limitations, we designed a lightweight 4D attention mechanism capable of precisely modeling the interannual evolution of dynamic postoperative processes such as liver regeneration and steatosis. This mechanism not only significantly reduces parameter count but also enhances dynamic perception of CRLM-specific biological processes.

In multimodal data fusion, dual challenges of spatiotemporal scale mismatch and redundant interference persist. Litjens et al. [19] identified that temporal-spatial disparities between clinical indicators (e.g., CEA) and imaging features (e.g., sinusoidal dilatation) may induce modality mismatch risks. For example, Tixier et al. [20] demonstrated that irinotecan therapy increases the coefficient of variation in CT texture features, exacerbating cross-modal feature alignment bias. Parmar et al. [21] compared early and late fusion strategies, concluding that neither can resolve the regulatory mechanisms of microenvironmental dynamics on imaging phenotypes. Consequently, developing causality-driven fusion models is critical. Our study proposes a Transformer-based cross-modal alignment strategy that jointly optimizes modality alignment loss and disentanglement loss functions, suppressing redundant feature interference and establishing chemotherapy response-related dynamic biomarkers. This approach provides a novel theoretical framework for multimodal dynamic modeling. These advancements not only offer new insights and methodologies for postoperative CRLM prognosis prediction but also lay a robust foundation for future research directions, driving further progress in the field.

## 3 Methods

The proposed CRLM prognostic prediction model, based on a multimodal spatiotemporal joint modeling framework, consists of three core modules: a 4D spatiotemporal attention mechanism, a clinical-imaging feature dynamic fusion module, and a multi-task prognostic prediction head, as illustrated in Figure 2.

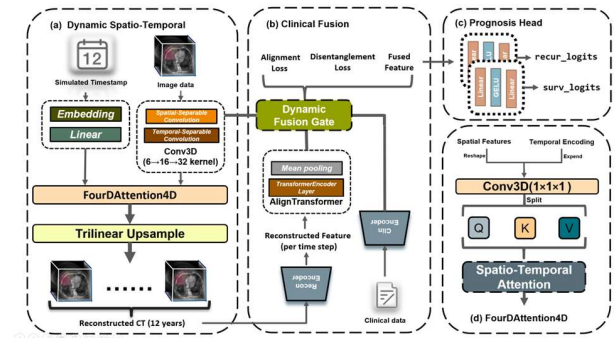

**Figure 2: CRLM Prognostic Prediction Framework Based on Multimodal Spatiotemporal Joint Modeling.**

### 3.1 4D Spatiotemporal Attention Mechanism

The 4D spatiotemporal attention mechanism serves as the core of our framework, enabling continuous modeling of postoperative intrahepatic dynamic evolution through virtual timestamp generation and spatiotemporal feature decoupling. As shown in Figure 3, this mechanism operates in four stages.

In the first stage, discrete timesteps $t\{\in 1,…,12\}$ (corresponding to postoperative years) are embedded into continuous vectors $E_t$ through an embedding layer. These vectors are then broadcast and spatially aligned with spatial feature maps to form spatiotemporally fused inputs $\widetilde{X}_t = [X_{ct}; E_t]$. In the second stage, spatiotemporal separable convolutions are employed to decouple spatial and temporal features. Specifically, a 3×3×3 convolutional kernel captures local spatial features such as tumor morphology and vascular topology, while a 3×1×1 convolutional kernel models microenvironmental evolution between adjacent timesteps. This design reduces parameter count by 41% compared to traditional 3D convolutions while retaining temporal modeling capability. In the third stage, 4D attention weights are computed. Spatiotemporal features and temporal encodings are concatenated along the channel dimension to generate query (Q), key (K), and value (V) vectors. The attention scores are then calculated as follows:

$$\text{Attention}(Q, K, V) = \text{softmax}\left(\frac{QK^T}{\sqrt{d_k}}\right)V \quad (1)$$

where $d_k$ represents the dimension of the key vectors. Based on the computed attention weights, features from each timestep are adaptively aggregated through weighted summation.

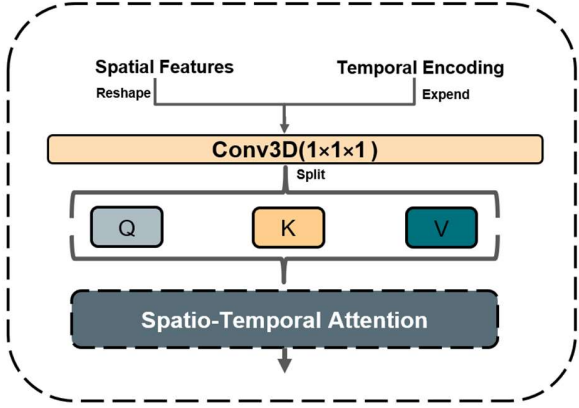

Figure 3: The 4D spatiotemporal attention mechanism.

Subsequently, transposed convolution is applied to restore spatial resolution, generating reconstructed feature maps that encode dynamic postoperative evolution. Finally, the aggregated features across all timesteps undergo deconvolution-based spatial upsampling, yielding high-resolution reconstructed images that reflect spatiotemporally integrated prognostic patterns.

## 3.2 Clinical-Imaging Feature Dynamic Fusion

This module aims to achieve efficient multimodal feature fusion, as illustrated in Figure 4. It comprises four submodules: clinical data encoding, reconstructed image feature extraction, spatiotemporal alignment, and dynamic gated fusion. A dual-path joint optimization strategy is introduced to constrain intermodal relationships, enabling the model to adaptively balance contributions from imaging and clinical features.

First, clinical data are mapped into a low-dimensional semantic space to filter redundant noise while retaining critical prognostic factors, yielding encoded clinical features $h_{clin}$. Next, for each timestep, reconstructed CT images undergo hierarchical feature extraction through a cascaded architecture, including local feature extraction, global feature compression, and cross-timestep dependency modeling using a Transformer encoder. The resulting time-aware global representations $h_{recon}$ are concatenated with spatial and temporal features $h_{img}$ extracted by the 4D attention mechanism. Finally, a learnable gating mechanism dynamically balances the contributions of these three feature types to generate the fused representation.

$$h_{fused} = \sigma(gate_0) \cdot h_{img} + \sigma(gate_1) \cdot h_{recon} + h_{clin} \quad (2)$$

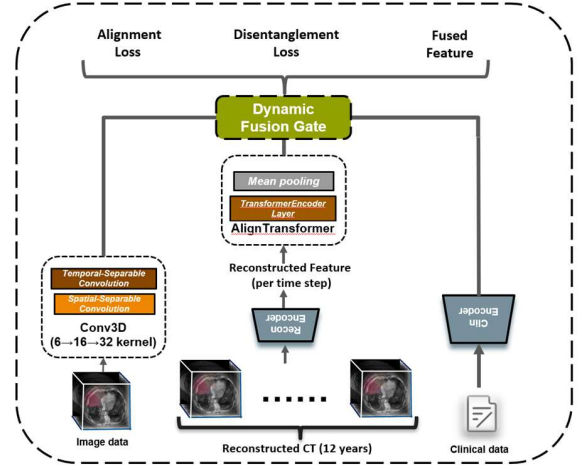

Figure 4: The Clinical-Imaging Feature Dynamic Fusion.

The module is trained using two complementary loss functions:
**Modality Alignment Loss.** Maximizes feature consistency between imaging and clinical features:

$$L_{align} = -\frac{1}{B}\sum_{i=1}^{B} \cos\left(h_{img}^{(i)}, h_{clin}^{(i)}\right) \quad (3)$$

where $\cos(\cdot)$ computes cosine similarity between feature vectors. The negative sign ensures that minimizing this loss equates to maximizing similarity.
**Modality Disentanglement Loss.** Minimizes covariance matrix differences to suppress multicollinearity:

$$L_{align} = \frac{\left\|\text{Cov}(H_{img}) - \text{Cov}(H_{clin})\right\|_F}{d^2} \quad (4)$$

Where $\|\cdot\|_F$ denotes the Frobenius norm. This constraint forces the model to isolate modality-specific information, enhancing feature discriminability.

For enhanced clarity, Figure 5 presents a comprehensive schematic diagram illustrating the implementation workflow of the module. This figure delineates the step-by-step process from the input data, encompassing imaging features $h_{img}$, clinical data $x_{clin}$, and reconstructed images $X_{recon}$, to the output fused features $h_{fused}$, along with the alignment loss $L_{align}$ and disentanglement loss $L_{dis}$. Through this schematic, we can gain an intuitive understanding of the interconnections between each step and the flow of data within the module.

```
Input: Raw imaging features h_img, clinical data x_clin, reconstructed images X_recon
Output: Fused features h_fused, loss L_align, L_dis

1. Encode Clinical Data:  h_clin ← ClinEncoder(x_clin)

2. Extract Reconstructed Features:
   for t=1 to T do
        h_recon^(t) ← ReconEncoder(X_recon^(t))
   end for
        h_recon ← Transformerer({h_recon^(t)}_{t=1}^T)

3. Compute Alignment Loss:  L_align ← −mean(cos(h_img, h_clin))

4. Compute Disentanglement Loss:  L_dis ← ||Cov(h_img) − Cov(,h_clin)||_F / d^2

5. Dynamic Fusion:  h_fused ← σ(gate_0)·h_img + σ(gate_1)·h_recon + h_clin
```

Figure 5: Clinical-Imaging Dynamic Fusion.

### 3.3 Multi-Task Prognostic Prediction Head

As illustrated in Figure 2c, this module predicts interannual recurrence probabilities and survival risks at 12 postoperative time points based on the fused features. Its core architecture consists of two parallel task branches: a recurrence classification head and a survival regression head. Both branches share the underlying fused features but are optimized through independent objective functions—cross-entropy loss for recurrence classification and mean squared error for survival regression—enabling multi-task collaborative learning.

The recurrence classification head employs a two-layer fully connected network (32→12 neurons) with GELU activation and a Sigmoid output layer to generate time-dependent recurrence probabilities. Similarly, the survival regression head utilizes an analogous structure (32→12 neurons) to output survival probabilities, with a Softmax function normalizing the probability distribution across the 12 months. Through this design, the model simultaneously captures shared information and task-specific characteristics between recurrence and survival tasks, providing comprehensive prognostic support for clinical decision-making.

## 4 Experiments

In this section, we validate the effectiveness of the proposed multimodal spatiotemporal joint modeling framework through comprehensive experiments. First, we conduct comparative experiments against the 4D spatiotemporal attention mechanism to evaluate its performance in CRLM prognosis prediction tasks. Subsequently, ablation studies are performed to verify the contributions of individual modules, supporting our hypothesis that the synergistic integration of the 4D spatiotemporal attention mechanism, cross-modal alignment and disentanglement strategies, and the multi-task prognostic prediction head significantly enhances the model's prediction accuracy and generalization capability.

### 4.1 Datasets and Evaluation Metrics

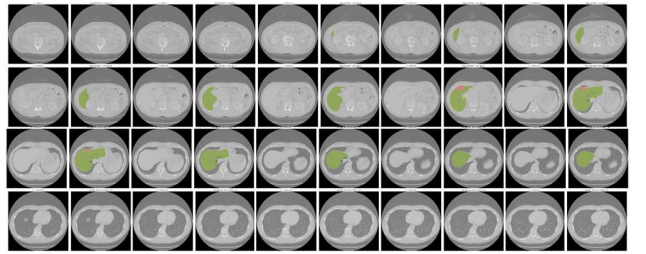

Figure 6: The data set used in the experiment.

**Datasets.** This study is based on a publicly available dataset from the Memorial Sloan Kettering Cancer Center (MSKCC), comprising multimodal data from 197 pathologically confirmed colorectal cancer liver metastasis (CRLM) patients [23]. Preoperative portal venous phase contrast-enhanced CT images for all patients were stored in DICOM format, covering the entire liver's 3D anatomical structure (resolution: 512×512×N). Segmentation labels (liver, future liver remnant, hepatic veins, portal veins, and tumor regions) were annotated by three imaging experts. Clinical data included age, sex, primary tumor TNM staging, distribution of liver metastases, serum CEA levels, and treatment regimens. The dataset was divided into training (157), validation (20), and test sets (20) in an 8:1:1 ratio.

**Data Preprocessing.** For image preprocessing, CT data were first enhanced using adaptive histogram equalization to improve tumor-liver parenchyma contrast, followed by downsampling to 128×128 resolution using third-order spline interpolation to reduce computational load. To address variability in the number of Z-axis slices, an anatomy-based adaptive cropping algorithm was applied: for cases with ≥40 slices, a 40-slice volume centered at the hepatic portal vein bifurcation was extracted; for cases with fewer slices, symmetric padding with -1024 HU was performed to ensure a uniform input size of 40 slices for the 3D convolutional network. Segmentation labels were refined using mathematical morphological closing operations to smooth jagged edges. Clinical data preprocessing included converting months to years. To enhance model generalization, a multi-dimensional data augmentation strategy was implemented: 1) Elastic deformation based on B-spline interpolation (control point grid: 7×7×7, maximum displacement: 5 mm) simulated intraoperative liver deformation and respiratory motion artifacts; 2) Gaussian noise ($\sigma \in [0, 0.1]$) and motion blur ($\sigma \in [0, 0.5]$) were added; 3) CT values were linearly mapped to the [-1, 1] range to eliminate density shifts caused by scanning protocol variations.

**Performance Metrics.** In this study, we employed temporal adjacency accuracy (TAA), mean squared error (MSE), and mean absolute error (MAE) to evaluate model performance. MSE measures the squared error between predicted and true values, while MAE assesses the absolute deviation between predicted and true values. TAA is defined as the proportion of samples where the absolute error between predicted and true recurrence/survival times is ≤1 year, directly reflecting the model's clinical utility. These metrics collectively provide a comprehensive assessment of the model's prediction accuracy and stability.

$$TP - Acc = \frac{1}{N}\sum_{i=1}^{N} \mathbb{I}\left(\left|t_{pred}^{(i)} - t_{true}^{(i)}\right| \leq 1\right) \quad (5)$$

where $\mathbb{I}(\cdot)$ is the indicator function, and $N$ is the number of samples. This metric directly reflects the model's practical utility in clinical decision support.

## 4.2 Implementation Details

To ensure fairness and reproducibility, all experiments were conducted under a unified hardware and software environment. The specific configuration is as follows: The hardware setup includes an NVIDIA RTX 4090 GPU (24GB VRAM). Due to the high memory demands of 3D data, the batch size was set to 1. The AdamW optimizer was employed with an initial learning rate of 1e-4, and a ReduceLROnPlateau scheduler (factor=0.5, patience=5) was used to dynamically adjust the learning rate based on validation loss, optimizing the training process. The loss function adopts a multi-task loss formulation, where both recurrence and survival losses are computed using cross-entropy loss, defined as:

$$L_{align} = 0.5L_{surv} + 0.3L_{recur} + 0.1L_{align} + 0.1L_{dis} \quad (6)$$

Here, $L_{surv}$ and $L_{recur}$ represent the cross-entropy losses for survival and recurrence prediction, while $L_{align}$ and $L_{dis}$ denote the modality alignment loss and modality disentanglement loss, respectively. Through weighted combination, the model simultaneously optimizes multiple tasks, ensuring synergistic interactions among them. Additionally, all experiments followed the same training and evaluation protocols to guarantee comparability and reliability of the results.

## 4.3 Comparative Experiments

To validate the effectiveness of the 4D spatiotemporal attention mechanism in capturing postoperative liver regeneration and microenvironmental dynamic evolution, we designed comparative experiments (see Table 1), comparing the proposed method with two baseline models: 3D CNN only (removing temporal dimension processing) and 3D CNN + LSTM (adding an LSTM module for temporal modeling). The 3D CNN only focuses on spatial features, while 3D CNN + LSTM introduces temporal modeling capabilities through LSTM but does not explicitly model spatiotemporal interactions. By comparing temporal adjacency accuracy (TAA), mean squared error (MSE), and mean absolute error (MAE), we comprehensively evaluated the impact of spatiotemporal modeling on prognosis prediction.

The experimental results demonstrate that the proposed method significantly outperforms 3D CNN only (TAA = 0.7000) and 3D CNN + LSTM (TAA = 0.8500) in temporal adjacency accuracy (TAA = 1.0000), while also achieving optimal performance in MSE (0.4250) and MAE (0.4250). In contrast, 3D CNN only (MSE = 2.5750, MAE = 1.1250) and 3D CNN + LSTM (MSE = 1.0625, MAE = 0.6375) exhibited inferior performance, indicating that relying solely on spatial features or simple temporal modeling is insufficient to fully capture postoperative dynamic changes. These results validate the effectiveness of the 4D spatiotemporal attention mechanism in spatiotemporal modeling, enabling more accurate prediction of postoperative prognosis for CRLM patients.

**Table 1: Comparative Experiments for Validating the Effectiveness of Spatiotemporal Modeling.**

| Model | Metrics | | |
|---|---|---|---|
|  | TAA | MSE | MAE |
| 3D CNN only (removing temporal dimension processing) | 0.7000 | 2.5750 | 1.1250 |
| 3D CNN + LSTM (adding an LSTM module for temporal modeling) | 0.8500 | 1.0625 | 0.6375 |
| 4D spatiotemporal attention mechanism | 1.0000 | 0.4250 | 0.4250 |

## 4.4 Ablation Studies

Our work incorporates three core innovations: the 4D spatiotemporal attention mechanism (A), cross-modal alignment operation (B), and modality disentanglement strategy (C). To analyze the contribution of each module, we conducted systematic ablation experiments. The baseline model employs only a 3D CNN for spatial feature extraction, without any spatiotemporal modeling or multimodal fusion strategies. We incrementally added each module and ultimately validated the performance of the complete model. The experimental results are presented in Table 2.

From Table 2, it can be observed that the baseline model performs poorly in temporal adjacency accuracy (TAA = 0.1750), mean squared error (MSE = 25.5875), and mean absolute error (MAE = 4.1875). This indicates that relying solely on spatial features is insufficient to capture postoperative liver regeneration and microenvironmental dynamic evolution, underscoring the necessity of spatiotemporal modeling and multimodal fusion. When the cross-modal alignment operation (3D-CNN + B) is introduced to the baseline model, performance improves significantly: TAA increases to 0.7000, MSE decreases to 2.5750, and MAE decreases to 1.1250. This improvement validates the importance of alignment operations in multimodal feature fusion, effectively mitigating scale mismatch between clinical data and imaging features.

Further incorporating the 4D spatiotemporal attention mechanism (A + B) enhances model performance: TAA reaches 0.9250, MSE decreases to 0.8625, and MAE decreases to 0.5875. The 4D spatiotemporal attention mechanism, through spatiotemporal separable convolution and virtual timestamp encoding, significantly strengthens the model's ability to capture postoperative dynamic changes. When the modality disentanglement strategy (A + B + C) is added to the cross-modal alignment operation, performance improves further: TAA reaches 0.9000, MSE decreases to 1.1250, and MAE decreases to 0.7500. This demonstrates that the disentanglement strategy effectively isolates modality-specific information, reducing interference from redundant features and enhancing the model's generalization capability.

The complete model (A + B + C), integrating the 4D spatiotemporal attention mechanism, cross-modal alignment operation, and modality disentanglement strategy, achieves optimal performance across all evaluation metrics. The experimental results indicate that the synergistic interaction of these modules significantly improves the model's prediction accuracy and robustness, providing robust technical support for postoperative prognosis prediction in CRLM.

Table 2: Different types of ablation experiments.

| Types | Metrics | | |
|---|---|---|---|
| | TAA | MSE | MAE |
| 3D-CNN | 0.1750 | 25.5875 | 4.1875 |
| 3D-CNN+Alignment | 0.7000 | 2.5750 | 1.1250 |
| 4D Spatiotemporal Attention + Alignment | 0.9250 | 0.8625 | 0.5875 |
| 3D-CNN + Alignment + Modality Disentanglement | 0.9000 | 1.1250 | 0.7500 |
| 4D Spatiotemporal Attention + Alignment + Modality Disentanglement | 1.0000 | 0.5625 | 0.4875 |

## 5 Conclusion

This study proposes a colorectal cancer liver metastasis (CRLM) prognosis prediction framework based on multimodal spatiotemporal joint modeling. By innovatively integrating a 4D spatiotemporal attention mechanism and a clinical-imaging dynamic alignment strategy, it achieves, for the first time, synergistic modeling of postoperative liver regeneration, microenvironment remodeling, and tumor heterogeneity. Experimental results demonstrate that the framework achieves a temporal adjacency accuracy (TAA) of 100% on the MSKCC dataset, with significantly improved performance, validating its advantages in capturing postoperative dynamic changes and multimodal feature fusion. The 4D spatiotemporal attention mechanism, through spatiotemporal separable convolution and virtual timestamp encoding, significantly reduces parameter count (by 41%) while enhancing the modeling capability of dynamic processes such as postoperative liver regeneration and steatosis. The cross-modal alignment and disentanglement strategies effectively mitigate scale mismatch and redundant interference between clinical data and imaging features, improving the effectiveness of multimodal fusion. The dynamic prognostic decision module generates personalized interannual recurrence risk heatmaps through temporal upsampling and a gated classification head, providing interpretable spatiotemporal decision support for clinical practice. This study establishes the first spatiotemporal modeling paradigm for postoperative dynamic monitoring of CRLM, and its framework can be extended to prognostic analysis of multi-cancer metastases, advancing precision surgery from "spatial resection" to "spatiotemporal cure."

## REFERENCES
[1] Ferlay J , Soerjomataram I , Dikshit R ,et al.Cancer incidence and mortality worldwide: Sources, methods and major patterns in GLOBOCAN 2012[J].International Journal of Cancer, 2015, 136(5):E359-E386.DOI:10.1002/ijc.29210.
[2] Freddie,Bray,Jacques,et al.Global cancer statistics 2018: GLOBOCAN estimates of incidence and mortality worldwide for 36 cancers in 185 countries.[J].CA: a cancer journal for clinicians, 2018.DOI:10.3322/caac.21492.
[3] Zarour L R , Anand S , Billingsley K G ,et al.Colorectal Cancer Liver Metastasis: Evolving Paradigms and Future Directions[J].Cmgh Cellular & Molecular Gastroenterology & Hepatology, 2017, 3(2):163-173.DOI:10.1016/j.jcmgh.2017.01.006.
[4] Filip S , Vymetalkova V , Petera J ,et al.Distant Metastasis in Colorectal Cancer Patients—Do We Have New Predicting Clinicopathological and Molecular Biomarkers? A Comprehensive Review[J].International Journal of Molecular Sciences, 2020, 21(15):5255.DOI:10.3390/ijms21155255.
[5] Rada M , Kapelanski-Lamoureux A , Petrillo S ,et al.Runt related transcription factor-1 plays a central role in vessel co-option of colorectal cancer liver metastases[J].Communications Biology, 2021, 4(1).DOI:10.1038/s42003-021-02481-8.
[6] Du F , Li X , Feng W ,et al.SOX13 promotes colorectal cancer metastasis by transactivating SNAI2 and c-MET[J].Oncogene[2025-02-28].DOI:10.1038/s41388-020-1233-4.
[7] Li Y , Jin-Si-Han E E M B K , Feng C ,et al.An evaluation model of hepatic steatosis based on CT value and serum uric acid/HDL cholesterol ratio can predict intrahepatic recurrence of colorectal cancer liver metastasis[J].International Journal of Clinical Oncology, 2024, 29(9).DOI:10.1007/s10147-024-02550-y.
[8] Lee H S , Kwon H W , Lim S B ,et al.FDG metabolic parameter-based models for predicting recurrence after upfront surgery in synchronous colorectal cancer liver metastasis[J].European Radiology, 2023, 33(3).DOI:10.1007/s00330-022-09141-3.
[9] Kim S K , Kim S Y , Kim C W ,et al.A prognostic index based on an eleven gene signature to predict systemic recurrences in colorectal cancer[J].Experimental & Molecular Medicine, 2019, 51(10):1-12.DOI:10.1038/s12276-019-0319-y.
[10] Li Y , Jin-Si-Han E E M B K , Feng C ,et al.An evaluation model of hepatic steatosis based on CT value and serum uric acid/HDL cholesterol ratio can predict intrahepatic recurrence of colorectal cancer liver metastasis[J].International Journal of Clinical Oncology, 2024, 29(9).DOI:10.1007/s10147-024-02550-y.
[11] Yu M , Yang C , Wang S ,et al.Plasma Methylated SEPT9 as a Novel Biomarker for Predicting Liver Metastasis in Colorectal Cancer[J].Molecular biotechnology, 2024(9):66.
[12] Brudvik K W , Jones R P , Giuliante F ,et al.RAS Mutation Clinical Risk Score to Predict Survival After Resection of Colorectal Liver Metastases.[J].Annals of Surgery, 2017:1.DOI:10.1097/SLA.0000000000002319.
[13] Andersen P K , Gill R D .Cox's Regression Model for Counting Processes: A Large Sample Study[J].Annals of Statistics, 1982, 10(4):1100-1120.DOI:10.1214/aos/1176345976.

[14] Geoff,Jones.Joint Models for Longitudinal and Time-to-Event Data: with Applications in R[J].Australian & New Zealand Journal of Statistics, 2013.
[15] Tan, J., Le, H., Deng, J. et al. Characterization of tumour heterogeneity through segmentation-free representation learning on multiplexed imaging data. Nat. Biomed. Eng (2025). https://doi.org/10.1038/s41551-024-01223-5
[16] Zhou X , Qian Y , Ling C ,et al.An integrated framework for prognosis prediction and drug response modeling in colorectal liver metastasis drug discovery[J].Journal of Translational Medicine, 2024, 22(1).DOI:10.1186/s12967-024-05127-5.
[17] Deng Y , Chen Q , Guo C ,et al.Comprehensive single-cell atlas of colorectal neuroendocrine tumors with liver metastases: unraveling tumor microenvironment heterogeneity between primary lesions and metastases[J].Molecular Cancer, 2025, 24(1):1-25.DOI:10.1186/s12943-025-02231-y.
[18] Bastiancich C, Snacel-Fazy E, Fernandez S, et al. Tailoring glioblastoma treatment based on longitudinal analysis of post-surgical tumor microenvironment[J]. Journal of experimental & clinical cancer research, 2024, 43(1): 311.
[19] Guo D , Lin M , Pei J ,et al.Tri-Modal Confluence withTemporal Dynamics forScene Graph Generation inOperating Rooms[C]//International Conference on Medical Image Computing and Computer-Assisted Intervention.Springer, Cham, 2024.DOI:10.1007/978-3-031-72089-5_67.
[20] Litjens G , Kooi T , Bejnordi B E ,et al.A Survey on Deep Learning in Medical Image Analysis[J].Medical Image Analysis, 2017, 42(9):60-88.DOI:10.1016/j.media.2017.07.005.
[21] Tixier F , Le Rest C C , Hatt M ,et al.Intratumor heterogeneity characterized by textural features on baseline 18F-FDG PET images predicts response to concomitant radiochemotherapy in esophageal cancer.[J].Journal of nuclear medicine : official publication, Society of Nuclear Medicine, 2011, 52(3):369-78.DOI:10.2967/jnumed.110.082404.
[22] Rios E , Parmar C , Jermoumi M ,et al.Robust Radiomics Feature Quantification Using Semiautomatic Volumetric Segmentation[J].Plos One, 2014, 41(6):452-452.DOI:10.1118/1.4889256.
[23] Simpson, A.L., Peoples, J., Creasy, J.M. et al. Preoperative CT and survival data for patients undergoing resection of colorectal liver metastases. Sci Data 11, 172 (2024). https://doi.org/10.1038/s41597-024-02981-2